\documentclass[letterpaper,twocolumn,prl,aps,superscriptaddress,amsmath,amssymb,floatfix]{revtex4-1}
\usepackage{mathptmx}
\usepackage[latin9]{inputenc}
\usepackage{color}
\usepackage{amsmath}
\usepackage{amssymb}
\usepackage{graphicx}
\usepackage{esint}
\usepackage[unicode=true,
 bookmarks=true,bookmarksnumbered=false,bookmarksopen=false,
 breaklinks=false,pdfborder={0 0 1},backref=false,colorlinks=true]
 {hyperref}
\hypersetup{
 linkcolor=magenta,urlcolor=blue,citecolor=blue,pdfstartview={FitH},hyperfootnotes=false}

\makeatletter

\usepackage{textcomp}
\usepackage{epstopdf}

\pdfpageheight\paperheight
\pdfpagewidth\paperwidth

\@ifundefined{textcolor}{}{%
 \definecolor{BLACK}{gray}{0}
 \definecolor{WHITE}{gray}{1}
 \definecolor{RED}{rgb}{1,0,0}
 \definecolor{GREEN}{rgb}{0,1,0}
 \definecolor{BLUE}{rgb}{0,0,1}
 \definecolor{CYAN}{cmyk}{1,0,0,0}
 \definecolor{MAGENTA}{cmyk}{0,1,0,0}
 \definecolor{YELLOW}{cmyk}{0,0,1,0}
}

\usepackage{xcolor}\usepackage{soul}
\definecolor{blue}{rgb}{0,0,1}
\definecolor{red}{rgb}{1,0,0}
\definecolor{green}{rgb}{0,1,0}

\usepackage{siunitx}
\usepackage{physics}
\usepackage{mathrsfs}

\usepackage{soul}

\makeatother

\begin{document}
\title{Chip-integrated Brillouin Saser Gyroscope}

\author{Wen-Qi Duan}
\affiliation{Laboratory of Quantum Information,
University of Science and Technology of China, Hefei 230026, China}
\affiliation{Anhui Province Key Laboratory of Quantum Network,
University of Science and Technology of China, Hefei 230026, China}

\author{Ming-Xuan Zhao}
\affiliation{Laboratory of Quantum Information,
University of Science and Technology of China, Hefei 230026, China}
\affiliation{Anhui Province Key Laboratory of Quantum Network,
University of Science and Technology of China, Hefei 230026, China}

\author{Jia-Qi Wang}
\affiliation{Laboratory of Quantum Information,
University of Science and Technology of China, Hefei 230026, China}
\affiliation{Anhui Province Key Laboratory of Quantum Network,
University of Science and Technology of China, Hefei 230026, China}

\author{Xin-Biao Xu}
\affiliation{Laboratory of Quantum Information,
University of Science and Technology of China, Hefei 230026, China}
\affiliation{Anhui Province Key Laboratory of Quantum Network,
University of Science and Technology of China, Hefei 230026, China}

\author{Luyan Sun}
\affiliation{Center for Quantum Information, Institute for Interdisciplinary Information
Sciences, Tsinghua University, Beijing 100084, China}
\affiliation{Hefei National Laboratory, Hefei 230088, China}

\author{Guang-Can Guo}
\affiliation{Laboratory of Quantum Information,
University of Science and Technology of China, Hefei 230026, China}
\affiliation{Anhui Province Key Laboratory of Quantum Network,
University of Science and Technology of China, Hefei 230026, China}
\affiliation{CAS Center For Excellence in Quantum Information and Quantum Physics,
University of Science and Technology of China, Hefei, Anhui 230026,
China}
\affiliation{Hefei National Laboratory, Hefei 230088, China}

\author{Ming Li}
\email{lmwin@ustc.edu.cn}
\affiliation{Laboratory of Quantum Information,
University of Science and Technology of China, Hefei 230026, China}
\affiliation{Anhui Province Key Laboratory of Quantum Network,
University of Science and Technology of China, Hefei 230026, China}
\affiliation{CAS Center For Excellence in Quantum Information and Quantum Physics,
University of Science and Technology of China, Hefei, Anhui 230026,
China}
\affiliation{Hefei National Laboratory, Hefei 230088, China}

\author{Chang-Ling Zou}
\email{clzou321@ustc.edu.cn}
\affiliation{Laboratory of Quantum Information,
University of Science and Technology of China, Hefei 230026, China}
\affiliation{Anhui Province Key Laboratory of Quantum Network,
University of Science and Technology of China, Hefei 230026, China}
\affiliation{CAS Center For Excellence in Quantum Information and Quantum Physics,
University of Science and Technology of China, Hefei, Anhui 230026,
China}
\affiliation{Hefei National Laboratory, Hefei 230088, China}

\date{\today}
\begin{abstract}
On-chip Brillouin laser gyroscopes harnessing opto-acoustic interaction are an emerging approach to detect rotation, due to their small footprint, excellent stability and low power consumption. However, previous implementations rely solely on optical readout, leaving the simultaneously generated saser (sound amplification by stimulated emission) undetected due to the lack of capability to access the acoustic output. Here, we propose a gyroscope based on saser detection using a suspension-free chip platform that supports low-loss confinement of both optical and acoustic modes. With experimental feasible parameter with optical and acoustic quality factors of $\sim 10^5$ and $5000$, respectively, sasers show significantly suppressed thermal and frequency noises, leading to gyroscope performance that outperforms its optical counterparts. We predict an angle random walk $\sim 0.1\,\mathrm{deg/\sqrt{h}}$ by saser gyroscope, while a conventional Brillouin laser gyroscope requires significantly higher pump power and optical quality factor to achieve comparable performance. Our work establishes the foundation for active phononic integrated circuits with Brillouin gain, opening avenues in inertial sensing, quantum transduction, and RF signal processing.
\end{abstract}
\maketitle

\noindent\textit{Introduction.- }Chip-integrated inertial sensors, including gyroscopes and accelerometers~\cite{Barbour2001,El-Sheimy2020}, are indispensable for modern navigation, robotics, consumer electronics, and autonomous systems. Conventional approaches rely on the inertial motion of mechanical structures, such as vibrating proof masses in micro-electro-mechanical systems~\cite{Yazdi1998,Shaeffer2013}  and trajectories of laser-cooled atoms in vacuum chambers~\cite{Kasevich1991,Gustavson1997,GarridoAlzar2019}, to detect rotation and acceleration. However, these mass-based approaches face fundamental constraints of limited stability and large footprint~\cite{Lee2022}. An alternative approach detects inertial motion through the dynamics of information carriers, including nuclear spin~\cite{Jarmola2021}, electromagnetic waves~\cite{Vali1976,Chow1985}, and acoustic waves~\cite{Oh2015,Fu2019}. Among these, optical approaches exploiting the Sagnac effect have been proven particularly powerful, enabling fiber-optic gyroscopes to achieve navigation-grade sensitivity with kilometers of optical paths~\cite{Nayak2011,Wang2022,Song2023}. For chip-scale implementation of optical gyroscope~\cite{Liang2017,Ai2025}, although limited by short propagation lengths, Brillouin laser gyroscopes have been proposed and demonstrated by leveraging ultra-narrow-linewidth Brillouin laser oscillations~\cite{Li2013,Gundavaraju2019,Lai2020}.

However, a fundamental symmetry between the photon and phonon in the pump-stimulated Brillouin interaction~\cite{Eggleton2019} has been overlooked. When the Brillouin threshold is exceeded, the system simultaneously generates both optical laser and saser (acoustic counterpart of laser) outputs~\cite{wang2023laser2}. While previous Brillouin gyroscopes exclusively detect the optical beatnote, the acoustic output has been neglected due to the incapability in detecting acoustic waves in conventional photonic chips. Recent breakthroughs in phononic integrated circuits~\cite{Fu2019,Mayor2021,Xu2022b,Xu2025a,Xu2025c} have fundamentally changed this landscape: acoustic waves at gigahertz frequencies can now be tightly confined and guided in crystalline thin-film platforms such as lithium niobate on sapphire, achieving acoustic quality factors approaching $10^4$ without suspended structures~\cite{Mayor2021,Wang2025}. Moreover, the excellent piezoelectric properties of lithium niobate enable highly efficient bidirectional conversion between acoustic waves and microwave signals through interdigital transducers (IDTs), providing direct electrical detection capability~\cite{Mayor2021,Xu2025b}. These advances raise a compelling question: could the acoustic output in Brillouin gyroscopes actually provide superior sensitivity when properly engineered?

\begin{figure*}[t]
\begin{centering}
\includegraphics[width=1\textwidth]{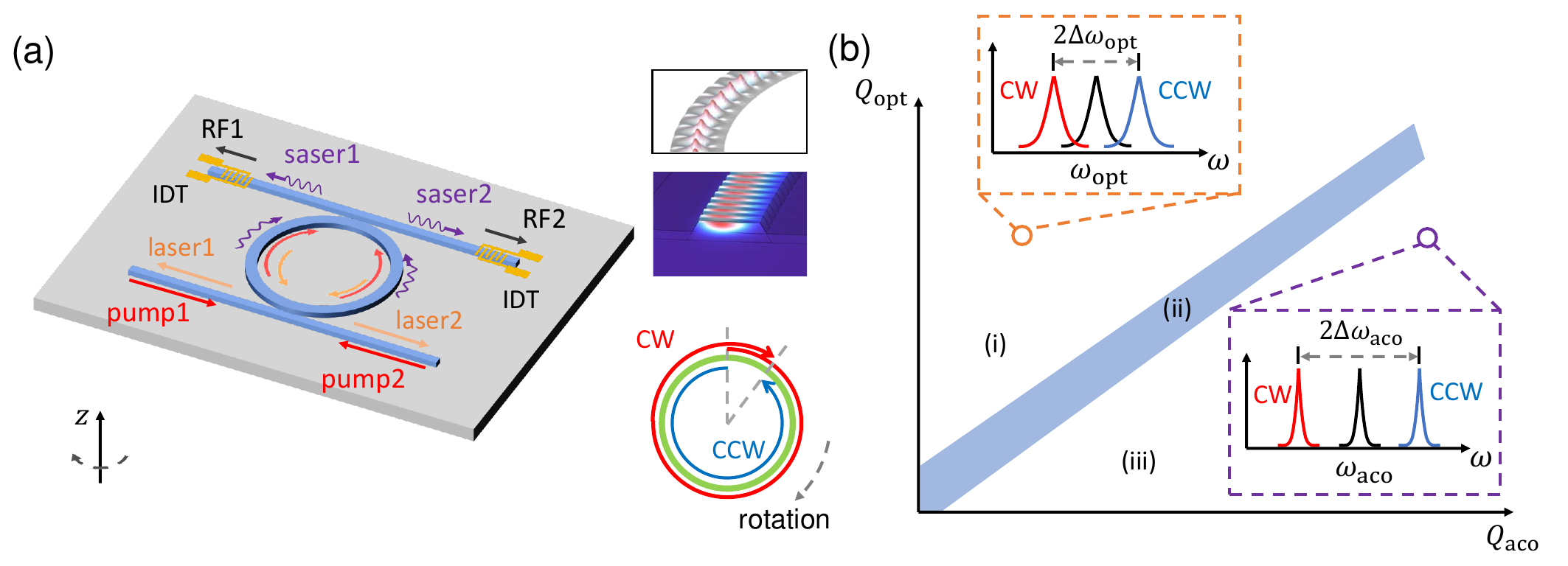}\end{centering}
\caption{Principle of an on-chip Brillouin saser gyro. (a) Schematic of the on-chip Brillouin saser gyroscope. Two counter-propagating optical beams are injected into the cavity to generate Brillouin sasers. Stokes photons are coupled out to the pumping waveguide, while phonons are extracted via the opposite waveguide and converted into microwave signals using an interdigital transducer (IDT). The insets display the simulated electric field distribution of the photonic mode and the displacement field of the phononic mode, along with an illustration of the Sagnac effect. Under rotation, the clockwise (CW, red arrow) and counter-clockwise (CCW, blue arrow) waves experience different round-trip path lengths, resulting in a resonance frequency splitting.  (b) Working regimes corresponding to different modal quality factors ($Q$). The laser and saser exhibit superior performance in regimes (i) and (iii), respectively, with a transition region in regime (ii). Since the acoustic wave vector is approximately twice of the optical wave vector, the acoustic Sagnac frequency shift is approximately twice of the optical wave, as depicted in the inset. }
\label{Fig1}
\end{figure*}

In this Letter, we investigate Brillouin gyroscopes operating in the high acoustic quality factor regime and demonstrate theoretically that saser-based detection can achieve dramatically enhanced sensitivity compared to conventional laser-based architectures. By analyzing the noise dynamics in hybrid photonic-phononic microring resonators, we reveal that pump frequency noise is distributed between optical and acoustic outputs according to their relative decay rates, with noises mainly being transferred to the mode with larger linewidth. Consequently, the system with a high acoustic quality factor will generate saser that effectively immune to pump noise while inheriting a factor-of-two Sagnac sensitivity enhancement. For achievable acoustic quality factor of $5000$ at room temperature, the direct saser detection can offer an angle random walk (ARW) $\sim0.1\,\mathrm{deg/\sqrt{\mathrm{h}}}$), requiring a relatively low optical quality factor and pump power. Beyond demonstrating a novel gyroscope architecture, this work establishes a potential Brillouin-gain enpowered phononic systems for advanced acoustic-domain signal processing and sensing in integrated chip platforms.

\smallskip{}
\noindent \textit{Principle.- }Figure~\ref{Fig1}(a) sketches the Brillouin microring cavity~\cite{Yang2023,Yang2024,Yang2025b,Xu2025b}, which can simultaneously support tight confinement of phononic and photonic modes (as shown by the insets) without suspended structures, for coherent photon-phonon coupling. The chip platform is based on a ``Zhengfu" architecture using thin-film lithium niobate on sapphire (LNOS) platform~\cite{Yang2025a}, allowing the suspension-free high-quality factor photonic and phononic modes with single-crystal materials. Compared to conventional photonic circuits on a chip, the Zhengfu architecture also allows the scalable phononic circuits~\cite{Xu2025a,Xu2025c} and high-efficiency electrical ports for efficient phononic excitation and collection~\cite{Yang2025a,Xu2025b}. Therefore, the microring supports coherent phonon-photon coupling when the phase-matching condition of backward Brillouin scattering is satisfied. In particular, around 9.0\,GHz phonon frequency, the Brillouin scattering works for 1550\,nm wavelength photons~\cite{Yang2023,Wang2025}.

For both acoustic and optical modes traveling along the microring, they will respond to the rotation of the chip due to the Sagnac effect. As shown by the inset of Fig.~\ref{Fig1}(a), traveling mode around the micoring has the round trip length modified by the rotation, and consequently the mode frequencies of the clockwise and counter-clockwise directions change with opposite signs. According to Sagnac effect, frequency shift of optical (acoustic) cavity mode is written as
\begin{equation}
        \Delta\omega_{\mathrm{opt(aco)}} = S_{\mathrm{opt(aco)}} \Omega = \dfrac{\omega_{\mathrm{opt(aco)}} R}{v_\mathrm{opt(aco)}}\Omega,
\end{equation}
where $S_{\mathrm{opt(aco)}}$ denotes the sensitivity and $\omega_{\mathrm{opt(aco)}}$ denotes optical (acoustic) resonance frequency, $\Omega$ denotes the gyro rotation rate, $R$ denotes cavity radius,  $v_\mathrm{opt(aco)}$ is light (acoustic) velocity in the waveguide. Here, both optical and acoustic frequency shift is proportional to cavity radius and inversely proportional to traveling velocity, indicating larger cavity and slower acoustic wave can further increase the Sagnac shift. Then, the CW and CCW modes have an opposite frequency shift as $\omega_{\mathrm{opt(aco)}}\pm\Delta\omega_{\mathrm{opt(aco)}}$. According to the phase matching condition of the backward Brillouin scattering, the acoustic wave-vector is approximately two times of that of the optical wave~\cite{Wang2025}, i.e., $\omega_{\mathrm{aco}}/v_{\mathrm{aco}}\approx2\omega_{\mathrm{opt}}/v_{\mathrm{opt}}$, thus $ S_{\mathrm{aco}} \approx 2 S_{\mathrm{opt}}$ means that the acoustic sensitivity is twice of the optical sensitivity.

\begin{figure*}
\includegraphics[width=1\textwidth]{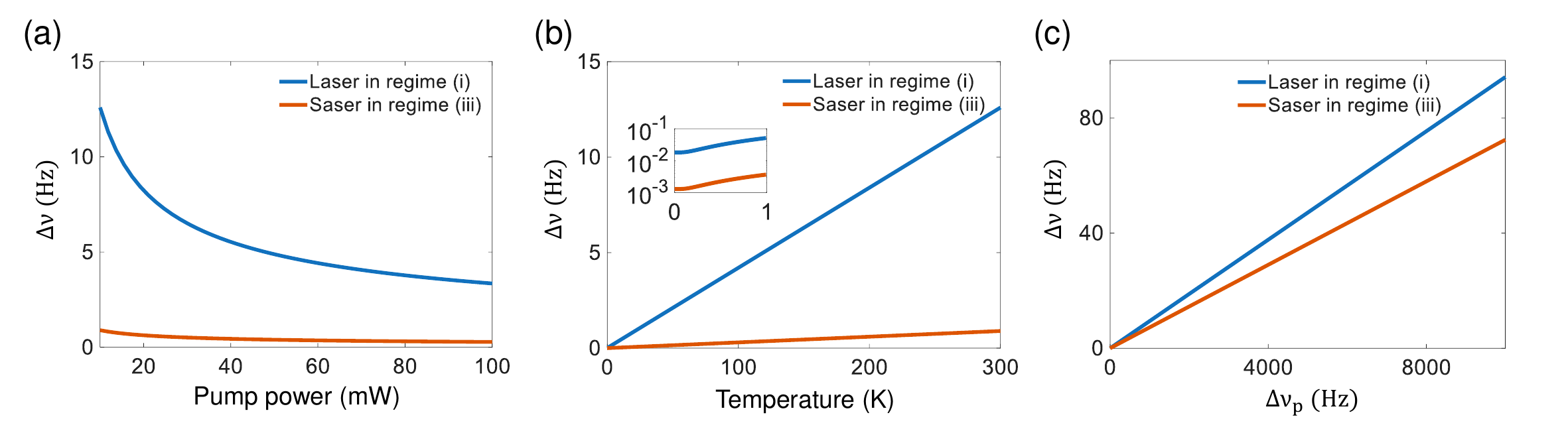}
\caption{Noise characteristics of Brillouin saser and laser. The optical quality factor is fixed at $Q_{\mathrm{opt}} = 10^7$, while the acoustic quality factors are $Q_{\mathrm{aco}} = 5 \times 10^1$ for the laser and $Q_{\mathrm{aco}} = 5 \times 10^3$ for the saser, respectively. (a) Dependence of linewidth on pump power ($P$) for the saser (orange curve) and laser (blue curve) under thermal noise. Both linewidths scale inversely to $P$, with the saser demonstrating a narrower linewidth than the laser across the measured range. (b) Temperature dependence of linewidths for both saser and laser under thermal noise. While the relationship is linear over a wide temperature range, nonlinear behavior emerges in the cryogenic regime (inset), consistent with the temperature dependence of thermal phonon populations.  (c) Pump transferred noise characteristics showing linewidth dependence on pump noise for saser and laser regimes. The output linewidth scales linearly with the pump linewidth, but is significantly suppressed through the Brillouin scattering process. }
\label{Fig2}
\end{figure*}

Although it is straightforward to derive the rotation rate of the chip by directly measure the change of amplitude or phase of the optical transmittance, the signal-to-noise ratio (SNR) of the passive optical gyroscope is limited due to the limited linewidths of the resonator modes and noise of the pump laser. Instead, it is widely adopted to introduce the gain to the optical resonators, which generates the laser signal and the $\omega_{\mathrm{opt(aco)}}$ can be directly detected through the beating signal of the CW and CCW laser outputs. The backward Brillouin interaction is a three-wave mixing process, as described by the Hamiltonian $\hbar g_0(a^\dagger_\mathrm{pump}a_\mathrm{aco}a_{\mathrm{opt}}+a_\mathrm{pump}a^\dagger_\mathrm{aco}a^\dagger_{\mathrm{opt}})$, where $g_0$ denotes the Brillouin phonon-photon interaction~\cite{Yang2024,Yang2025b} and $a_{\mathrm{pump,opt,aco}}$ are the bosonic operators of the involved modes. Under a strong pump to optical mode ($a_\mathrm{pump}$), the Brillouin interaction can provide the gain to both optical signal mode ($a_\mathrm{opt}$) and acoustic mode ($a_\mathrm{aco}$), with the effective amplitude decay rates of the optical (acoustic) mode being modified to
\begin{equation}
\kappa_{\mathrm{opt(aco),eff}}=(1-C)\kappa_{\mathrm{opt(aco)}},
\end{equation}
where $C=N_\mathrm{pump}g^2_0/\kappa_\mathrm{opt}\kappa_\mathrm{aco}$ is the photon-phonon coupling cooperativity stimulated by the intracavity pump photon number $N_\mathrm{pump}$, $\kappa_\mathrm{opt}$ and $(\kappa_\mathrm{aco})$ are amplitude decay rates of bare optical and acoustic mode ($N_\mathrm{pump}=0$), respectively. When the gain exceeds certain threshold, i.e., the above express is not valid as $C\geq1$ or $\kappa_{\mathrm{opt(aco),eff}}$ becomes unphysical, we realize highly compact active Brillouin gyroscope with simultaneous laser and saser outputs. 

As shown in Fig.~\ref{Fig1}(a), by simultaneously pumping from CW and CCW direction, we can either collect the output saser signal from the bus phonon waveguide through the two IDTs and measure the microwave beating signal, or collect the output laser signal from the bus photonic waveguide and measure the optical beating signal. We derive the beat frequency of both Brillouin saser (acosutic detection) and laser (optical detection) from steady state solution of the system as follows
\begin{equation}
    {\nu}_{\mathrm{beat}}= S_\mathrm{eff}\left|\dfrac{\Omega}{2\pi}\right|,
\end{equation}
with an effective sensitivity $S_\mathrm{eff}= ({\kappa_{\mathrm{aco}}} S_{\mathrm{opt}}+{\kappa_{\mathrm{opt}}}S_{\mathrm{aco}})/({\kappa_{\mathrm{aco}}+\kappa_\mathrm{opt}})$. Therefore, measurements of beatnote in all conditions are directly proportional to gyro rotation rate. Considering in the high acoustic $Q$ regime ($\kappa_\mathrm{aco}\ll\kappa_{\mathrm{opt}}$), the response of the acoustic wave dominates with $  {S}_{\mathrm{eff}}\approx S_{\mathrm{aco}} \approx 2S_{\mathrm{opt}}$, while in the reverse case,  $  {S}_{\mathrm{eff}}\approx S_{\mathrm{opt}}$ is determined by the optical Sagnac effect.

\begin{figure*}[t]
\begin{centering}
\includegraphics[width=1.0\textwidth]{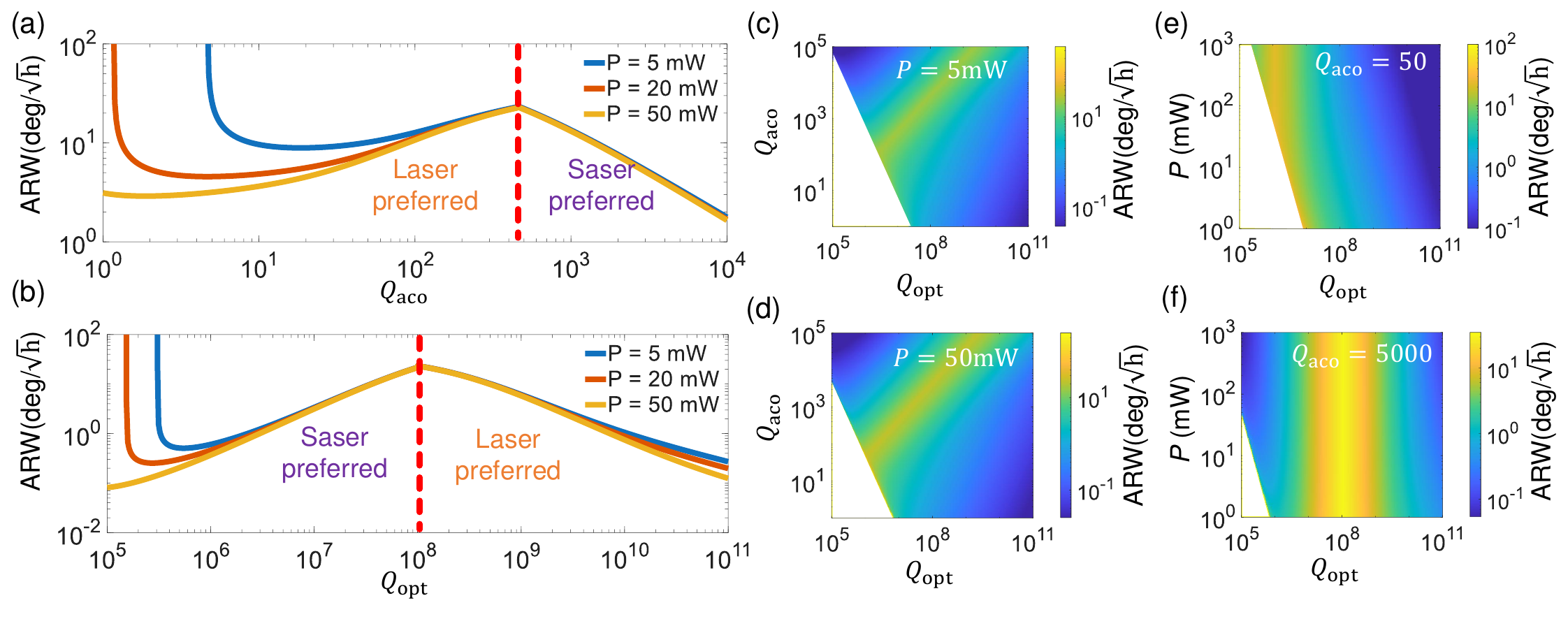}
\end{centering}
\caption{Gyroscope performance characterization. Pump linewidth is fixed at 1 kHz. (a,b) ARW dependence on modal quality factors with (a) fixed optical quality factor $Q_{\mathrm{opt}}=1\times10^7$ and (b) fixed acoustic quality factor $Q_{\mathrm{aco}}=5\times10^3$. Brillouin saser and laser gyroscopes exhibit superior performance in distinct quality factor regimes. Increasing pump power ($P$) reduces ARW in both cases, necessitating higher $Q_{\mathrm{opt}}$ or $Q_{\mathrm{aco}}$ values to establish the absolute advantage of either saser-based or laser-based gyroscopes.  (c,d) ARW as a function of modal quality factors with $P$ fixed at (c) 5 mW and (d) 50 mW. The blank regions correspond to parameter spaces where the Brillouin threshold exceeds 5 mW and 50 mW, respectively. Within the working regions, saser and laser regimes are separated by a distinct cross-over band.  (e,f) ARW dependence on $Q_{\mathrm{opt}}$ and $P$ with $Q_{\mathrm{aco}}$ fixed at (e) $Q_{\mathrm{aco}}=50$ and (f) $Q_{\mathrm{aco}}=5000$. }
\label{Fig3}
\end{figure*}

\smallskip{}
\noindent\textit{Laser and saser frequency noise.- } Although there are simultaneously generated saser and laser signal when pump above the threshold, the noise performance of the acoustic and optical output ports are significantly different. Intuitively, as shown in Fig.~\ref{Fig1}(b), there are three working regimes of the system under Brillouin gain above the threshold: (i) the lasing regime, (ii) cross-over regime, and (iii) saser regime. Most conventional Brillouin lasing belongs to the regime (i). In this regime, the acoustic linewidth is large, since the acoustic phonons in the structure can not be well confined or due to the strong phonon attenuation in silica or other non-crystalline materials. Consequently, the acoustic mode could be treated as a reservior and the laser frequency is determined by the optical resonance. When there is fluctuation in pump frequencies, the optical lasing frequency is less sensitive, while the frequency fluctuations are mostly transferred to the acoustic output frequency due to the energy conservation relation.

To quantitatively study the noise performance of the output laser and saser, we numerically investigated the output signal linewidths with realistic pump laser frequency fluctuations. Additionally, the thermal equilibrium phonon excitation number ($n_\mathrm{th}=(\mathrm{exp}(\hbar\omega_\mathrm{aco}/k_BT)-1)^{-1}$, with $T$ being the temperature and $k_B$ is the Boltzmann constant) will also affect the linewidth. Typically, thermal photon number is close to $n_\mathrm{th}\approx0$ in room temperature, while thermal phonon number is about $n_\mathrm{th}=700$ on the LNOS platform. By decomposing three mode operators into steady-state solutions and perturbation terms, noise dynamics can be well described by linearized Langevin equations, and we obtain the corresponding approximate output laser and saser frequency linewidths
\begin{eqnarray}
    \Delta\nu_{\mathrm{laser}} = (n_\mathrm{th}+1)\Delta\nu_{0}+\left( \dfrac{\kappa_{\mathrm{opt}}}{\kappa_{\mathrm{opt}}+\kappa_{\mathrm{aco}}}\right)^2\Delta\nu_\mathrm{p},\\
   \Delta\nu_{\mathrm{saser}} = (n_\mathrm{th}+1)\Delta\nu_{0}+\left( \dfrac{\kappa_{\mathrm{aco}}}{\kappa_{\mathrm{opt}}+\kappa_{\mathrm{aco}}}\right)^2\Delta\nu_\mathrm{p},\label{Eq:linewidth}
\end{eqnarray}
in regimes (i) and (iii), respectively. Here,
\begin{equation}
\Delta\nu_0=\dfrac{\kappa_{\mathrm{opt}}/\kappa_{\mathrm{aco}}}{\left( \kappa_{\mathrm{opt}}/\kappa_{\mathrm{aco}}+1\right)^2}\dfrac{\kappa_{\mathrm{opt}}}{4\pi N_\mathrm{aco}}
\end{equation} is the intrinsic laser linewidth and  $N_\mathrm{aco}$ denotes intracavity phonon numbers. The first terms of both $\Delta\nu_{\mathrm{laser}}$ and $\Delta\nu_{\mathrm{saser}}$ are equal as it originates from the same thermal phonon noise, while the second term of  $\Delta\nu_{\mathrm{laser}}$ and $\Delta\nu_{\mathrm{saser}}$ is from the transferred pump frequency noise $\Delta\nu_\mathrm{p}$.  Due to the mode pulling effect, the pump phase is distributed into to modes according to the relative modal decay rate, i.e., the pump transferred frequency noise is suppressed for laser in regime (i) and for saser in regime (iii), respectively.

In Fig.~\ref{Fig2}, we numerically investigated the different contributions to the output frequency noises with realistic experimental parameters~\cite{Yang2025b}, assuming both acoustic and optical modes are critically coupled, with $(Q_{\mathrm{aco}},\,Q_{\mathrm{opt}})=(50,\,10^7)$ and $(5000,\, 10^7)$ for regime (i) and regime (iii), respectively. This corresponds to a reduction of $\kappa_\mathrm{aco}/2\pi$ from $90\,\mathrm{MHz}$ to $0.9\,\mathrm{MHz}$ given a fixed $\kappa_\mathrm{opt}/2\pi=10\,\mathrm{MHz}$. Figures~\ref{Fig2}(a) and (b) show the relationships between linewidths of saser (laser) and pump power ($P$) and temperature, by neglecting the pump noise ($\Delta \nu_\mathrm{p}=0$). We should note that, the saser and laser share the same linewidth when $\Delta \nu_\mathrm{p}=0$. Comparing the output linewidths in all cases, the performance is significantly improved in regime (iii) compared with regime (i), by only enhancing the $Q_\mathrm{aco}$. Actually, any methods, enhancing intracavity phonon (photon) numbers or reducing thermal phonon numbers, are beneficial for a narrower saser (laser) linewidth by enhancing the contrast between the coherent and noise excitations. We should note that for even cryogenic temperatures, as shown by the inset of Fig.~\ref{Fig2}(b), the linewidth saturates at the intrinsic linewidth, showing a $\nu_0$ of 18\,mHz in regime (i) and 1.3\,mHz in regime (iii), corresponding to the enhancement of $N_\mathrm{aco}$ from $4.7\times10^{7}$ to $5.8\times10^{8}$ by reducing the acoustic dissipation.

In Fig.~\ref{Fig2}(c), the pump frequency fluctuation is considered. In regime (i), we have $\kappa_{\mathrm{aco}}\ll\kappa_{\mathrm{opt}}$, while we have $\kappa_{\mathrm{opt}}\ll\kappa_{\mathrm{aco}}$ in regime (iii), saser or laser linewidth is strongly suppressed from pump noise, as shown in Figure~\ref{Fig2}(c). The Eq.~(\ref{Eq:linewidth}) indicates that a larger contrast of the acoustic and optical dissipation rate will be beneficial, and for the parameters consider here the contrast in two regime are close thus showing similar behavior with around two-orders of magnitude of pump frequency noise suppression. It is anticipated that a higher $Q_\mathrm{aco}$ will be harmful to a conventional Brillouin laser gyroscope, as the pump transferred noise is increased, even though higher pump laser power leads to a suppressed $\nu_0$.

\smallskip{}
\noindent \textit{ARW.- }
To investigate the ultimate limit of the Brillouin gyroscope performance, we focus on the linewidths of laser and saser outputs by introducing the ARW as
\begin{equation}
    \zeta = \sqrt{\dfrac{4\pi\times\mathrm{min}\{\Delta\nu_{\mathrm{laser}},\Delta\nu_{\mathrm{saser}}\}}{ S_{\mathrm{eff}}^2}}
\end{equation}
by selectively detecting the optical or acoustic signal for best performance. This metric quantifies the gyroscope's sensitivity assuming ideal detection conditions and emphasizing the device's inherent noise properties over external factors in practical experimental setups, such as sampling frequency or detector bandwidth.

We compare the performances of room-temperature Brillouin gyroscope by varying acoustic and optical quality factors at different pump powers when $\Delta\nu_{\mathrm{p}}=1\, \mathrm{kHz}$, as illustrated in Figs.~\ref{Fig3}. For a fixed $Q_{\mathrm{opt}}=10^7$ [Fig.~\ref{Fig3}(a)], the $\zeta$ exhibits two distinct operating regimes. In the strong acoustic dissipation regime ($Q_{\mathrm{aco}}\sim10$), corresponding to the lasing preferred operation with suppressed pump transferred noise, $\zeta$ decreases with $P$ by enhancing the coherent intracavity excitations, while there is an optimal $Q_\mathrm{aco}$ due to the competition between the thermal and pump transferred noises. When the acoustic dissipation significantly reduced ($Q_{\mathrm{aco}}\gg10^3$), the saser detection is preferred. $\zeta$ decreases monotonously with $P$ and $Q_\mathrm{aco}$ by suppressing the contributions from both thermal and pump transferred noises. In particular, a moderate $Q_\mathrm{aco}=5000$ can outperform all laser-based operations when $P\leq{20}\,\mathrm{mW}$. Figure~\ref{Fig3}(b) shows a complementary analysis for fixed $Q_\mathrm{aco}=5000$, revealing a similar behavior. The saser regime dominates at moderate $Q_{\mathrm{opt}} \sim 2 \times 10^5$, while the lasing regime show advantages by requiring exceptionally high $Q_{\mathrm{opt}} > 10^{10}$ which is extremely challenging in current integrated platforms~\cite{Liu2022}. When $P$ increases, the optimal $Q_{\mathrm{opt}}$ decreases for better suppression of pump noise.

Figure~\ref{Fig3}(c) shows the landscape of the $\zeta$ with a given $P=5\,\mathrm{mW}$, with the blank region indicating parameters where the Brillouin gain is below threshold ($C < 1$).  The two operating regimes are clearly displayed, implying that the current LNOS platform~\cite{Yang2025a,Yang2025b} is well located in the saser operation regime. With a higher $P=50\, \mathrm{mW}$ [Fig.~\ref{Fig3}(d)], the sub-threshold area naturally shrinks. In the operating regimes, the sensitivity is quite similar to that in Fig.~\ref{Fig3}(c), indicating the increased $P$ helps little to the improvement of gyroscope.

Figure~\ref{Fig3}(e) illustrates the $\zeta$ of conventional Brillouin laser gyroscope with $Q_{\mathrm{aco}}=50$. Due to the high acoustic dissipation, Brillouin gyroscope is only preferred in the laser regime, showing that higher $Q_{\mathrm{opt}}$ yields better performance. However, an excellent performance area of saser emerges in the high acoustic ($Q_{\mathrm{aco}}=5000$) and low optical quality factors ($10^5\sim 10^6$) regime, as shown in Fig.~\ref{Fig3}(f). With a $P=200\, \mathrm{mW}$, Brillouin saser gyro exhibits a considerable {sensitivity ($\zeta=0.085\,\mathrm{deg/\sqrt{h}}$),} which is only achievable when $Q_{\mathrm{opt}}>10^{10}$ for laser preferred operation regime [Fig.~\ref{Fig3}(f)], or requires 100-times stronger pump power with optimistic $Q_\mathrm{opt}=10^8$ in conventional laser gyroscopes [Fig.~\ref{Fig3}(e)].

\smallskip{}
\noindent\textit{Conclusion.- } A gyroscope mechanism based on saser is proposed and numerically investigated. In conventional Brillouin laser gyroscope approaches, a high acoustic dissipation, which common in photonic platforms, benefits the gyroscope measurement by suppressing the pump frequency noise. However, we found that by only increasing the acoustic quality factor to $10^3$ level in the suspension-free Zhengfu architecture~\cite{Yang2025a,Yang2025b,Xu2025b,Wang2025}, a new operation regime emerges, showing unprecedented capability for direct phonon detection, high signal-to-noise ratio, while requiring lower pump laser power and lower optical mode quality factor. With experimental feasible optical quality factors ($10^5\sim10^6$), we predict that a device working in the saser regime with experimentally feasible acoustic quality factor ($Q_{\mathrm{aco}}=5000$) can significantly reduce the required pump power. In addition, an {ARW of $0.085(0.51)\,\mathrm{deg/\sqrt{h}}$} is predicted with saser operation at $P=200(5)\, \mathrm{mW}$, while similar performance is only possible with impractical ultrahigh optical quality factor ($>10^{10}$) which also introduces other restrictions~\cite{Matsko2018}. Beyond immediate applications in chip-scale inertial navigation, this work establishes the active phononic integrated circuits~\cite{Xu2025a,Xu2025c} with Brillouin-interaction-induced gain, opening new directions in precision measurement, quantum transduction, and RF signal processing.

\smallskip{}
\begin{acknowledgments}
This work was funded by the National Natural Science Foundation of China (Grants No.~92265210, 92165209, 92365301, 12374361, 11925404, 123B2068, 12104441, 12061131011). This work is also supported by the Fundamental Research Funds for the Central Universities, the USTC Research Funds of the Double First-Class Initiative, the supercomputing system in the Supercomputing Center of USTC the USTC Center for Micro and Nanoscale Research and Fabrication.
\end{acknowledgments}

\end{document}